\documentclass[a4paper,twocolumn,11pt]{quantumarticle}
\pdfoutput=1
\usepackage[utf8]{inputenc}
\usepackage[english]{babel}
\usepackage[T1]{fontenc}
\usepackage{amsmath}
\usepackage{hyperref}

\usepackage{tikz}
\usepackage{lipsum}

\usepackage{soul}
\usepackage[title]{appendix}%

\begin{document}
\title{A Vlasov-Bohm approach to Quantum Mechanics for statistical systems}

\author{Pedro Luis Grande}
\affiliation{Instituto de Física, Universidade Federal do Rio Grande do Sul, Av. Bento Gon\c{c}alves 9500, Porto Alegre, CP 15051, RS, Brazil}
\email{grande@if.ufrgs.br}
\author{Raul Carlos Fadanelli}
\affiliation{Instituto de Física, Universidade Federal do Rio Grande do Sul, Av. Bento Gon\c{c}alves 9500, Porto Alegre, CP 15051, RS, Brazil}
\email{raul@if.ufrgs.br}

\author{Maarten Vos}
\affiliation{Department of Materials Physics, the Australian National University, ANU, Canberra, 0200, ACT, Australia}
\email{Maarten.Vos@anu.edu.au}

\begin{abstract}
  Quantum mechanics is the most successful theory to describe microscopic phenomena. It was derived in different ways over the past 100 years by Heisenberg, Schr\"{o}dinger, and Feynman. At the same time, other interpretations have been suggested, including the Bohm-De Broglie interpretation and the so-called Bohmian mechanics. Here, we show that Bohmian mechanics, which utilizes the concept of the Bohm quantum potential, can also serve as a starting point for quantizing classical non-relativistic systems. By incorporating the Bohm quantum potential into the Vlasov framework, we obtain a mean-field theory that captures the corpuscular nature of matter, in agreement with quantum mechanics within the Random Phase Approximation (RPA).
\end{abstract}  


\section{Introduction}
The formal structure of quantum mechanics (QM) was established through diverse and seemingly unrelated developments by some of the most influential physicists of the early 20th century. In 1924, Heisenberg, along with Born and Jordan, introduced matrix mechanics, a formulation grounded in observable quantities and algebraic manipulation \cite{BornJordan:1925,Heisenberg:1925}. Shortly thereafter, in 1925, Schr\"{o}dinger proposed wave mechanics \cite{Schrodinger:1926}, based on a partial differential equation now bearing his name. Despite their mathematical differences, Schr\"{o}dinger and Heisenberg’s formulations were shown to be mathematically equivalent and to describe the same physical phenomena. Later, in the 1940s, Feynman \cite{Feynman:1948} introduced an alternative viewpoint through the path integral formalism, providing a new perspective grounded in the action principle of classical mechanics. These multiple representations, though different in interpretation and methodology, are all expressions of the same underlying theory of quantum mechanics.

Among these diverse formulations, Madelung \cite{Madelung1927322} introduced a hydrodynamic representation of quantum mechanics by expressing the complex wave function in polar form. This approach led to what became known as hydrodynamic quantum mechanics \cite{Hirschfelder1974,BONILLALICEA2021127171}, which later inspired Bohm to develop a deterministic interpretation of quantum theory. Initially developed by David Bohm in 1952 \cite{Bohm:1952}, Bohmian mechanics introduces a quantum potential derived from the Schr\"{o}dinger equation, which yields deterministic particle trajectories guided by the wave function. Unlike the Copenhagen interpretation, which emphasizes wavefunction collapse and probabilistic outcomes, Bohmian mechanics provides a trajectory-based framework where uncertainty arises not from indeterminism but from our lack of knowledge about initial conditions. This formulation has sparked considerable philosophical debate, but it has also proven to be a consistent and powerful interpretative tool, particularly in understanding quantum phenomena from a classical-like perspective \cite{Durrbook}. For more details on Bohmian mechanics, please see \cite{BohmianMechanicsWebsite}. 
Similarly, E. Nelson \cite{Nelson1966} proposed stochastic mechanics, in which particles undergo a Brownian motion–like process characterized by a diffusion coefficient but without friction. This framework resembles the quantum dynamics governed by the Bohm potential.

In conventional presentations, Bohmian mechanics is derived as an alternative reading of Schr\"{o}dinger’s wave mechanics \cite{Messiah:1967}. However, in this work, we propose reversing this logical sequence. Rather than viewing Bohmian mechanics as a secondary construct, we investigate whether its foundational elements—namely, the quantum potential and trajectory-based dynamics—can serve as a primary framework for quantization. By employing basic physical assumptions, such as the de Broglie postulate and the Heisenberg uncertainty principle, we reconstruct the structure of the quantum potential in a manner that allows us to directly quantize classical statistical systems, without recourse to the full machinery of wave mechanics. This nontraditional approach provides an intuitive bridge between classical statistical mechanics and quantum behavior, and may be particularly useful in systems where wavefunction-based formulations are cumbersome or less insightful.

This work aims to expand the interpretational and methodological toolkit of quantum mechanics by treating Bohmian mechanics not merely as an alternative viewpoint, but as a potential foundational structure from which quantum dynamics can be directly inferred at the non-relativistic level—similarly to Nelson’s stochastic mechanics \cite{Nelson1966}, which has recently been applied to machine learning based on stochastic processes \cite{WPaul2023}. We present a simple route to quantizing a classical statistical ensemble, avoiding the use of the Wigner distribution. A more general statistical formulation of quantum mechanics is discussed elsewhere \cite{Rundle2019,MORANDI2022128223}.

The paper is structured as follows. In the first part, we explore one-dimensional stationary systems and demonstrate that their quantum properties can be inferred from classical statistical ensembles by invoking the Bohm quantum potential. We compare these quantized results with exact solutions of the Schr\"{o}dinger equation to validate the approach. This provides a foundational test for the methodology, helping to establish confidence in its broader applicability.
In the second part, we extend the framework to collective classical systems described by the Vlasov equation, which is commonly used in plasma physics \cite{Vlasov_1968,nicholson:plasma,Bohm_1949,Bohm_1949B,Pines_1961} and fluid dynamics to represent statistical mixtures without collisions. The Vlasov equation, being inherently classical, does not admit particle-like solutions in its standard form. However, upon linearization and with the addition of the Bohm quantum potential, the corpuscular nature of quantum excitations emerges naturally. The resulting dispersion relations align with those derived from traditional quantum mechanics and the random phase approximation (RPA), suggesting that quantum corrections can be systematically incorporated into a classical formalism.

Finally, in Appendix A, we present a conceptual justification for the structure of the Bohm quantum potential. Although our reasoning is heuristic and based on qualitative arguments—sometimes referred to as "hand-waving"—we demonstrate that these arguments are surprisingly compelling in reconstructing the correct quantum dynamics. Appendix A also serves to underline the physical intuition that underpins this nontraditional route to quantization.

\section{Bohmian Quantum Potential and the Nonlinear Schr\"{o}dinger Equation}

In the Bohmian formulation of quantum mechanics, classical dynamics can be extended into the quantum regime through the introduction of the \emph{quantum potential} $Q(\vec{r})$, defined as \cite{Bohm:1952}
\begin{equation}
Q(\vec{r}) = -\frac{\hbar^2}{2 m A} \nabla^2 A,
\label{Q}
\end{equation}
where $m$ is the particle mass and $A$ is the amplitude of the wave function $\psi = A e^{iS/\hbar}$. For a system of independent particles, $A$ reduces to $\sqrt{\rho}$, with $\rho$ being the probability density. When combined with the continuity equation for the phase $S$, this form of $Q$ allows one to recover the full Schr\"{o}dinger equation via the Hamilton-Jacobi formalism~\cite{Messiah:1967}. A derivation of the Bohm quantum potential independent of the Schr\"{o}dinger equation is provided in Appendix A, based on heuristic physical arguments.

For one-dimensional systems and stationary states, a direct connection between the classical probability density and the total potential can be obtained from the continuity equation or, more simply, by invoking the classical relation that the probability of finding a particle at a given position is inversely proportional to its velocity. This yields
\begin{equation}
A^2 = \rho(x) = \frac{v_0}{\mathrm{v}(x)} = \frac{K \sqrt{E}}{\sqrt{ E - V_T(x) }},
\label{rho1d}
\end{equation}
where $v_0$ and $K$ are normalization constants (corresponding to the density at infinity for scattering states ), $E$ is the total energy, and $V_T(x)$ is the total potential. In Bohmian mechanics, $V_T$ is composed of the classical potential $V(x)$ and the quantum potential $Q(x)$, i.e., $V_T(x) = V(x) + Q(x)$.

By inverting Eq.~(\ref{rho1d}), the quantum potential can be expressed as
\begin{equation}
Q(x) = E \left( 1 - \frac{K^2}{A^4} \right) - V(x).
\label{Q2}
\end{equation}
This expression can also be derived using the microcanonical ensemble, yielding the same result and thereby reinforcing the statistical consistency of the formulation.

Substituting Eq.~(\ref{Q2}) into Eq.~(\ref{Q}) leads to a nonlinear differential equation for the amplitude $A(x)$, which takes the form of a nonlinear Schr\"{o}dinger equation (NLSE):
\begin{equation}
-\frac{\hbar^2}{2m} \frac{\text{d}^2 A}{\text{d}x^2} = \left[ E \left( 1 - \frac{K^2}{A^4} \right) - V(x) \right] A.
\label{nlse}
\end{equation}
Equation~(\ref{nlse}) is an important result of this work. It allows the construction of quantum-mechanical solutions from classical statistical quantities, while preserving equivalence with the original linear Schr\"{o}dinger equation. The effective potential entering the classical equations of motion is thus modified as
\begin{equation}
V_T(x) = V(x) + Q(x) = E \left( 1 - \frac{K^2}{A^4} \right).
\label{Vtot}
\end{equation}

\begin{figure}[ht]
\begin{center}
\includegraphics[width=6cm,angle = -90]{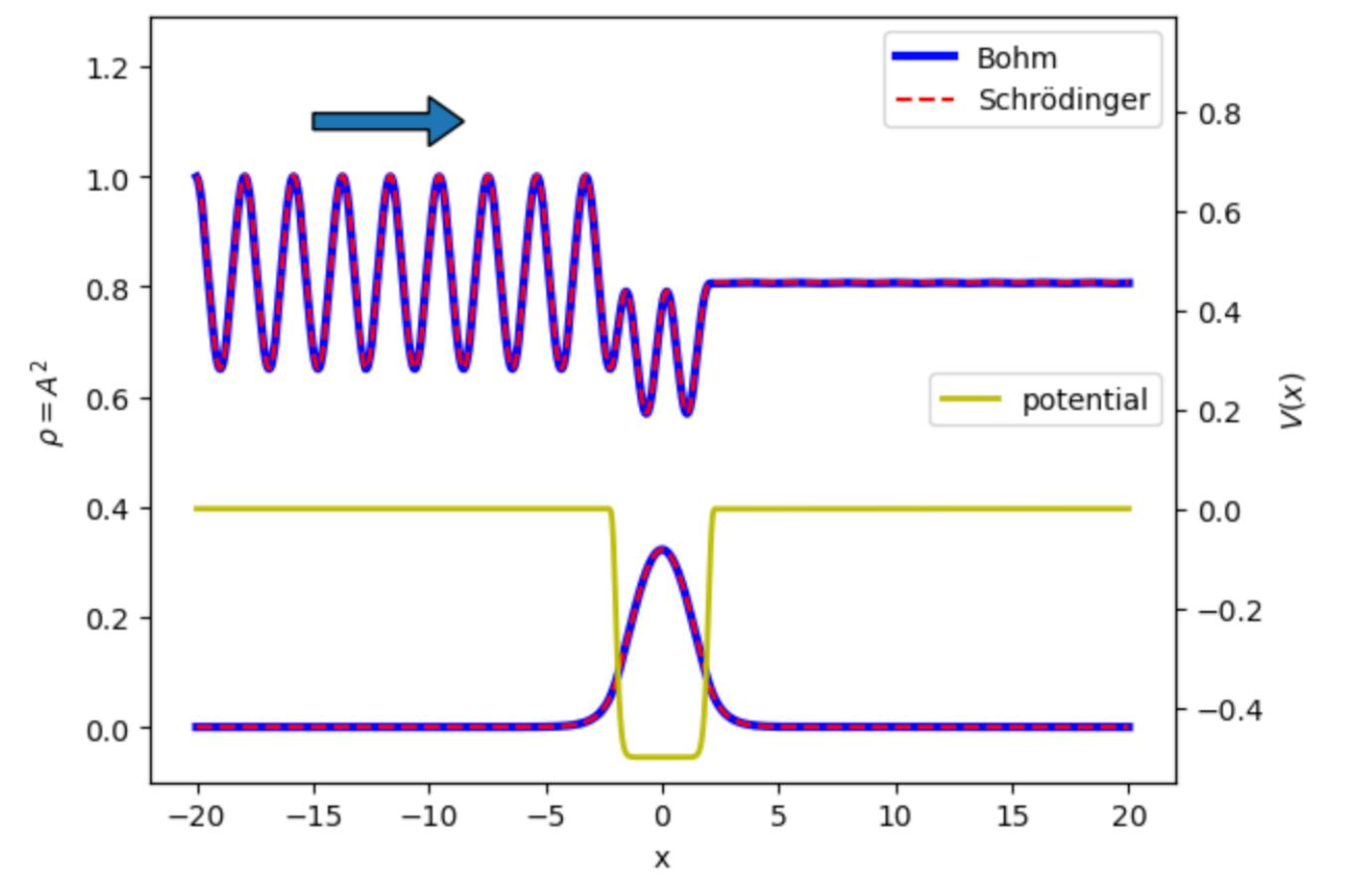}
\caption{Ground-state and scattering solutions of the Bohm and Schr\"{o}dinger equations for a potential well with depth \( V_0 = -0.5 \). The ground-state energy is \( E = -0.36 \), and the scattering solution corresponds to a wave number \( k = 1.5 \) (\( E = 1.125 \)) from left to right. See Eq.~(\ref{nlse}) for the definition of the nonlinear Schr\"{o}dinger equation underlying the Bohmian formulation.
}
\label{fig1}
\end{center}
\end{figure}

We emphasize that the solutions of Eq.~(\ref{nlse}) are in complete agreement with the stationary solutions of the time-independent Schr\"{o}dinger equation. Figure~\ref{fig1} illustrates this equivalence for both bound ($E<0$) and scattering ($E>0$) states in a potential $V(x)=V_0\exp\left ((x/L)^{n}\right )$.  Using  $n=16$, it resembles a finite square well of width $2L$ and height $|V_0|$.
Here we used $L=2$ and $V_0 = -0.5$. For bound states, integrable solutions for $A(x)$ exist only when $E$ corresponds to a true eigenvalue of the Schr\"{o}dinger problem, confirming the quantization condition. In the case of scattering states, the constant $K$ acquires a physical interpretation as proportional to the incoming wave density.

This formalism has also been extended to three-dimensional systems. Notably, recent studies have employed it to evaluate electronic stopping power in a free-electron gas (FEG), demonstrating its relevance for condensed matter systems and transport theory~\cite{Bohmpaper}.

\section{Quantum (Bohm) Vlasov linearized solution}
\label{Secvlaov}
The Vlasov equation, a fundamental tool in classical statistical mechanics, governs the evolution of the distribution function \( f(\vec{r}, \vec{\mathrm{v}}, t) \), which represents the density of particles in phase space interacting via a self-consistent mean field. In plasma physics, it is widely used to describe collective phenomena such as Langmuir waves.  Assuming small perturbations around a homogenous and stationary equilibrium $f_0(\vec{\mathrm{v}})$, we have:
\begin{eqnarray}
f(\vec{r}, \vec{\mathrm{v}}, t) &\approx& f_0(\vec{\mathrm{v}})\rho(\vec r,t) \\ 
 &\approx& f_0(\vec{\mathrm{v}}) + f_1(\vec {r}, \vec{\mathrm{v}}, t), 
\end{eqnarray}
with \( f_1 \ll f_0 \).
We now include the Bohm quantum potential, $Q(\vec{r})$, in the potential term of the linearized Vlasov equation. For details on the linearization procedure and further definitions, see Appendix B:

\begin{equation}
    q\phi(\vec r) \to q\phi(\vec r) + Q(\vec r)
\end{equation}
where 
\begin{equation}
    Q(\vec r) = -\frac{\hbar^2}{4 m}\frac{\nabla^2\rho}{\rho} - \frac{1}{2} \frac{\nabla\rho \cdot \nabla \rho}{{\rho}^2},
\end{equation}
and therefore
\begin{equation}
    Q_1(\vec r) = -\frac{\hbar^2}{4 m}\frac{\nabla^2 f_1}{f_0} 
\end{equation}
and the corresponding Fourier component :
\begin{equation}
    \hat Q_1(\vec k,\omega) = \frac{\hbar^2}{4 m}\frac{k^2}{f_0} \hat f_1(\vec k,\omega) 
\end{equation}
By proceeding $ q \hat \phi_1(\vec k,\omega) \to  q \hat \phi_1(\vec k,\omega) +  \hat Q_1(\vec k,\omega) $ in the linearized Vlasov equation solution Eq. (\ref{fromVlasov}) we have:
\begin{eqnarray}
    {\hat f}_1 (\vec k, \omega) = \frac{1}{m}k^2 &f_0&(\vec {\mathrm{v}}) \frac{1}{(\omega-\vec k \cdot \vec {\mathrm{v}})^2  } \\ \nonumber &\times& ( q \hat \phi_1(\vec k,\omega) +   \hat Q_1(\vec k,\omega)) 
\end{eqnarray}

\begin{eqnarray}
    {\hat f}_1 (\vec k, \omega) =\frac{1}{m} k^2 &f_0&(\vec {\mathrm{v}}) \frac{1}{(\omega-\vec k \cdot \vec {\mathrm{v}})^2  }\\ \nonumber &\times& ( q \hat \phi_1(\vec k,\omega)    +\frac{\hbar^2}{4 m}\frac{k^2}{f_0} f_1(\vec k,\omega)) 
\end{eqnarray}

\begin{equation}
    {\hat f}_1 (\vec k, \omega) = \frac{q}{m}\frac{ k^2 f_0(\vec {\mathrm{v}})    }{  (\omega-\vec k \cdot \vec {\mathrm{v}})^2 - \frac{\hbar^2}{4 m^2} k^4 } \hat \phi_1(\vec k,\omega) 
\end{equation}

\begin{eqnarray}
    &{\hat f}_1& (\vec k, \omega) =  \frac{q}{\hbar}f_0(\vec {\mathrm{v}}) \hat \phi_1(\vec k,\omega) \\ \nonumber &\times&\left (\frac{ 1    }{  (\omega-\vec k \cdot \vec {\mathrm{v}}) - \frac{\hbar}{2 m} k^2 }  - \frac{ 1  }{  (\omega-\vec k \cdot \vec {\mathrm{v}}) + \frac{\hbar}{2 m} k^2 }  \right)
\end{eqnarray}
and then 
\begin{eqnarray}
 &{\hat \rho}_1& (\vec k, \omega) = \int d^3 v  \frac{q}{\hbar}f_0(\vec {\mathrm{v}}) \hat \phi_1(\vec k,\omega)\\ \nonumber &\times& \left (\frac{ 1    }{  (\omega-\vec k \cdot \vec {\mathrm{v}}) - \frac{\hbar}{2 m} k^2 }  - \frac{ 1  }{  (\omega-\vec k \cdot \vec {\mathrm{v}}) + \frac{\hbar}{2 m} k^2 }  \right)  
\end{eqnarray}
Using the Poisson equation, Eq. (\ref{poisson}), we finally get

\begin{equation}
 \epsilon(\vec k, \omega) = 1 + \frac{q^2}{\epsilon_0 k^2}\int d^3 v \frac{ f_0(\hbar \vec k/m + \vec {\mathrm{v}}) -f_0(\vec {\mathrm{v}})   }{ \hbar \omega- \hbar\vec k \cdot \vec {\mathrm{v}} - \frac{\hbar^2}{2 m} k^2 } 
 \label{epsqm}
\end{equation}
which is the exact formula for $\epsilon(\vec k, \omega)$ for quantum mechanics for RPA, from where we can get the so-called Lindhard and Kaneko's dielectric functions \cite{Archubi2022,Kaneko2025}.

The Vlasov equation describes a continuum theory, where the dynamics of the system depends solely on the charge-to-mass ratio, $q/m$. Consequently, if electrons were hypothetically divided into two entities, each possessing half the mass and half the charge, the theory would remain entirely unaffected. Even fundamental quantities such as the plasma frequency, $\omega_p^2 = q^2 n / \epsilon_0 m$, would remain invariant under this transformation, as $q^2$ would decrease by a factor of four, the particle density $n$ would double, and $m$ would be halved, leaving $\omega_p$ unchanged.

This invariance underscores a key limitation of the Vlasov framework: it cannot account for inherently discrete particle effects. In the thermodynamic limit, the Vlasov equation holds true as the collision term decreases in importance\cite{Braun1977}. Features such as the Bethe ridge, which emerge from binary scattering kinematics and depend explicitly on the particle mass, are beyond the scope of a purely continuum description. 
However, the introduction of the Bohm potential fundamentally alters this picture. Although the quantum potential involves the probability density, the presence of $\hbar$ and $m$ introduces an explicit dependence on the mass of individual particles.

This breaks the charge-to-mass symmetry of the Vlasov framework, rendering the theory sensitive to the granular, corpuscular nature of matter. Under this modified theory, the hypothetical splitting of electrons into two fractional entities would indeed alter the dynamics, as both the quantum potential and the resulting wave-like behavior depend on the mass of the constituents. In this sense, the modified equation ceases to be a continuum theory; it becomes a quasi-particle theory, where discrete effects emerge naturally, and features like the Bethe ridge can be recovered.

Thus, the addition of the Bohm quantum potential  Eq.(\ref{Q}) to the linearized Vlasov equation results in the familiar RPA equation as derived in more conventional quantum physics by  Bohm and Pines \cite{PhysRev.92.609}.

\section{Numerical Example}
We illustrate now the classical and quantum limit for the case of a Gaussian occupancy: $ f_0(\vec k)= c e^{-k^2/\kappa^2}$.  Within atomic physics, such occupancy is often used as a first-order approximation of the momentum density of the core-level wavefunction, and $\kappa$ is chosen to achieve the best agreement with, e.g., a Hartree-Fock calculation. Kaneko \cite{Kaneko1989,Archubi2017a} determined analytical expressions for the RPA dielectric function for Gaussian occupancy). Within plasma physics \cite{Archubi2022}, a Gaussian occupancy for electrons is the consequence of a Maxwell-Boltzmann distribution of a plasma at temperature $T$: $ f_0(\vec k)= c e^{-E/k_b T}$ with $E$ the kinetic energy of the electrons ($\frac{\hbar^2 k^2}{2m}$) and $k_b$ the Boltzmann constant.  The relation between $\kappa$ and $T$ is thus for a Maxwell-Boltzmann distribution: $\kappa=\sqrt{2 m k_b T/\hbar^2}$.

The dielectric constant based on the RPA is then obtained by evaluating Eq. (\ref{epsqm}) for this occupation function. For most applications, the most useful representation of the dielectric function is the loss function $\rm{Im}\left[-1/\epsilon( k, \omega )\right]$. To avoid having to deal with singularities in the loss function, it is often convenient to add a small imaginary component to $\omega$: $\omega_c=\omega+i\Gamma$. 
For the RPA solution with Gaussian occupancy, one can then write for $\varepsilon ( \vec k, \omega)$, using the reduced variables $u=(\omega_c)/(\hbar k \kappa)$ and $z'=k/2\kappa$ \cite{Kaneko2025}:  .
\begin{equation}
    \epsilon(k, \omega)= 1+\frac{ \kappa^2 }{i \hbar^2 k^3}\left(w(u+z')-w(u-z')\right),\label{eq:rpa_solution}
\end{equation}  
where $w(z)$ is the Faddeeva function:
\begin{equation}
    w(z)=\frac{i}{\pi}\int_{-\infty}^\infty\frac{e^{-t^2}}{z-t}dt%=\frac{2iz}{\pi}\int_{0}^\infty\frac{e^{-t^2}}{z^2-t^2}dt.
\end{equation}.

The classical ($\hbar \to 0$, or small $k$) solution can be obtained from Eq.(\ref{eq:rpa_solution}) directly, as $z'$ is then small and thus $(w(u+z')-w(u-z')\approx  2z' w'(u)$.  For the derivative of the Faddeeva function, one can write:
\begin{equation}
    w^\prime(z) = \frac{2 i}{\sqrt \pi} - 2 z w(z)
\end{equation}.

\begin{figure}[ht]
\begin{center}
\includegraphics[width=12cm,]{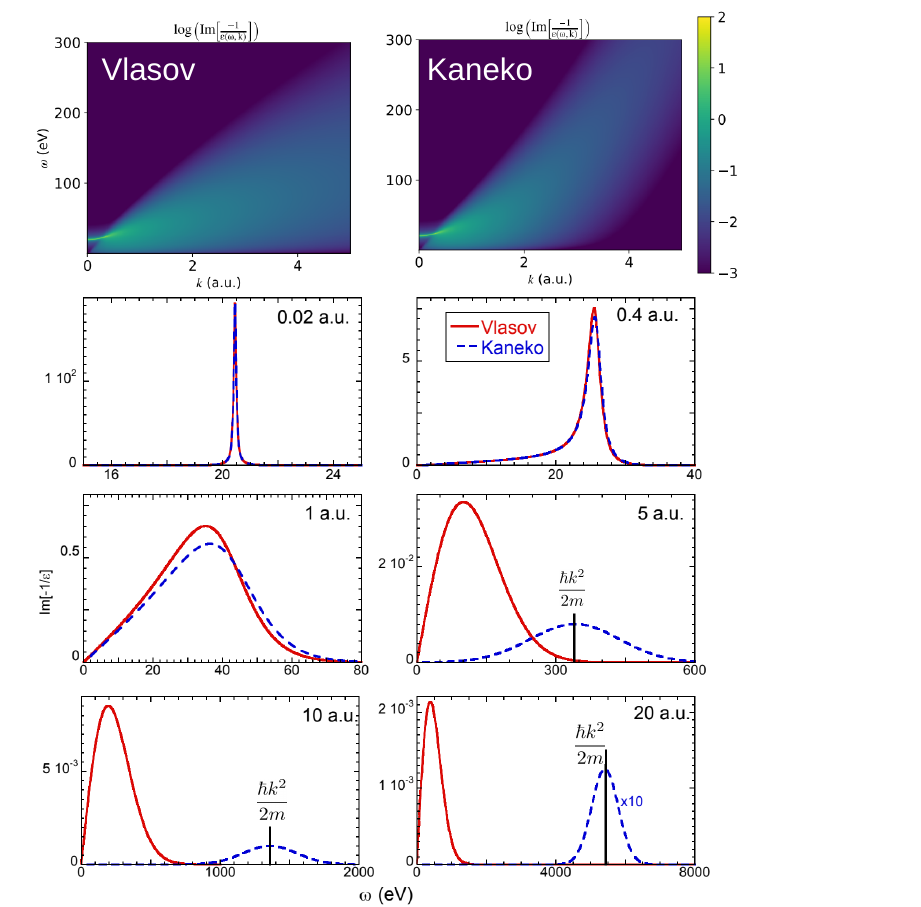}
\caption{False color plot of the loss function for the Vlasov and Kaneko dielectric function with $\kappa=1$ a.u. (top panels) The lower panels show the loss function for the indicated $k$ values. Only the Kaneko loss function is centered at $\frac{\hbar k^2}{2m}$ (`Bethe ridge') for large $k$ values.  }
\label{fig2}
\end{center}
\end{figure}

In Fig. \ref{fig2} we plot the loss function at several $k$ values 
for both the classical (`Vlasov') and quantum (`Kaneko') case for $\kappa=1$ a.u. and $\Gamma= 0.1$ eV. For $k<0.5$ a.u. both solutions are virtually identical. For large $k$-values, only the quantum solution peaks at the Bethe ridge, i.e., $\hbar^2 k^2/2m$. It is worth noting that both solutions adhere to the Bethe and F sum rules at all $k$ values. Archubi and Arista discussed the absence of the Bethe ridge in the Vlasov framework \cite{Archubi2022}, who showed that it leads to modest yet significant modifications in the stopping power calculations for ions in plasma.

\section{Conclusion}

In this work, we present an alternative and nontraditional approach to quantum mechanics by employing the Bohmian formulation as a foundational framework, rather than as an interpretation derived from the Schr\"{o}dinger equation. This approach enables the quantization of systems described by classical statistical ensembles through the introduction of the Bohm quantum potential. The resulting nonlinear Schr\"{o}dinger equation reproduces the exact stationary solutions of the standard quantum formalism.

Beyond its theoretical interest, this formulation offers practical advantages. It provides a bridge between classical and quantum descriptions, making it particularly suitable for systems governed by statistical mixtures, such as those found in plasma physics \cite{Haas:2011}. Furthermore, the pedagogical value of this approach is notable. By emphasizing the role of trajectories and classical concepts, it offers an intuitive introduction to non-relativistic quantum mechanics within the Bohmian framework.

Nevertheless, important open questions remain. Chief among them is the challenge of formulating an entirely consistent relativistic extension of Bohmian mechanics. Addressing this issue would represent a significant step toward unifying deterministic interpretations with the demands of relativistic quantum field theory.

\section{Acknowledgement}
We want to thank many colleagues from the Physics Department of UFRGS for stimulating discussions.
This study was financed in part by the Coordenação de Aperfeiçoamento de Pessoal de Nível Superior - Brazil (CAPES) - Finance Code 001, by FINEP, by Conselho Nacional de Desenvolvimento Científico e Tecnológico (CNPq) process number 403722/2023-3, and by Instituto Nacional de Engenharia de Superfícies (INES)  project number 465423/2014-0.

\begin{appendices}
\section{Heuristic Derivation of the Bohm Quantum Potential}

The Bohm quantum potential can be heuristically motivated by analyzing the spreading of a probability distribution for a free particle. In particular, we consider the case of an initial Gaussian wave packet and examine how its time evolution, governed by the uncertainty principle, gives rise to a quantum-like behavior even within a classical statistical framework.

The uncertainty principle relates the standard deviations of position and momentum as
\begin{equation}
\sigma_p \, \sigma_x(0) \ge \frac{\hbar}{2},
\label{uncertainty-principle}
\end{equation}
where $\sigma_x(0)$ is the initial spatial width of the wave packet, and $\sigma_p$ is the corresponding width in momentum space.

By solving the time-dependent equation for an initial Gaussian wave packet \cite{griffiths}, one obtains the familiar time-dependent spreading of a free-particle wave packet:
\begin{equation}
\sigma_x^2(t) = \sigma_x^2(0) + \left( \frac{\sigma_p}{m} t \right)^2.
\label{qm-spreading}
\end{equation}

Interestingly, this same spreading can be reproduced using a purely classical ensemble of non-interacting particles subject to an effective time-dependent repulsive potential of the form
\begin{equation}
V_{\text{spread}}(x, t) = -\frac{1}{2} s^2(t) x^2,
\label{spread-potential}
\end{equation}
where $s(t)$ is a time-dependent function representing the strength of the repulsive interaction. Such a potential leads to a linear force that decreases over time as the distribution broadens. For a Gaussian ensemble, this gives rise to the spreading:
\begin{equation}
\sigma_x(t) = \sigma_x(0) \cosh \left( \int_0^t \mathrm{d}t' \, s(t') \right).
\end{equation}

To match Eq.~(\ref{qm-spreading}), one can choose $s(t)$ as
\begin{equation}
s(t) = \frac{\sigma_p}{m \sigma_x(0)} \left[ 1 + \left( \frac{\sigma_p}{m \sigma_x(0)} t \right)^2 \right]^{-1/2}.
\label{s-choice}
\end{equation}

So far, these results are purely classical, provided that $\sigma_x(0)$ and $\sigma_p$ are treated as independent parameters. However, by invoking the uncertainty principle [Eq.~(\ref{uncertainty-principle})], we introduce quantum content into the dynamics.

The Bohm quantum potential $Q(x)$ should satisfy a few fundamental requirements:
\begin{enumerate}
    \item It must depend solely on the probability density $\rho(x)$.
    \item Its Fourier transform must scale as $k^2$ in the limit of small perturbations, consistent with known quantum dispersion relations (e.g., from the linearized Vlasov equation with quantum corrections).
    \item For a Gaussian distribution, it must reproduce the effective spreading potential $V_{\text{spread}}(x,t)$ given by Eq.~(\ref{spread-potential}).
\end{enumerate}

These constraints determine the Bohm potential (up to a constant) as
\begin{equation}
Q(x) = -c_\hbar \frac{ \partial_x^2 \sqrt{\rho(x)} }{ \sqrt{\rho(x)} },
\label{Q-bohm}
\end{equation}
where $c_\hbar$ is a constant to be determined. Matching this expression with Eq.~(\ref{spread-potential}) in the case of a Gaussian distribution yields
\begin{equation}
Q(x) = -\frac{1}{2} s^2(t) x^2.
\end{equation}

By evaluating the right-hand side of Eq.~(\ref{Q-bohm}) for a Gaussian distribution $\rho(x) \propto \exp[-x^2 / (2\sigma_x^2(t))]$, one finds:
\begin{equation}
Q(x) = \frac{c_\hbar}{\sigma_x^4(t)} \left( x^2 - \sigma_x^2(t) \right).
\end{equation}

Comparing coefficients with the effective potential in Eq.~(\ref{spread-potential}) fixes the value of $c_\hbar$:
\begin{equation}
c_\hbar = \frac{1}{2} s^2(t) \sigma_x^4(t).
\end{equation}

Now substituting $s(t)$ from Eq.~(\ref{s-choice}) and using the uncertainty principle from Eq.~(\ref{uncertainty-principle}), we find
\begin{equation}
c_\hbar = \frac{\hbar^2}{2m},
\end{equation}
which recovers the standard Bohm quantum potential in one dimension:
\begin{equation}
Q(x) = -\frac{\hbar^2}{2m} \frac{\partial_x^2 \sqrt{\rho(x)}}{ \sqrt{\rho(x)} }
\end{equation}
and the diffusion coefficient proposed by Nelson\cite{Nelson1966}.

Although this derivation is not rigorous, especially since it is not possible to derive a physical law, it provides a compelling physical motivation for the Bohm potential based on the classical-statistical behavior of free-particle ensembles and the minimal addition of quantum uncertainty. It also highlights that the quantum potential emerges naturally from the requirement that the spreading of the ensemble obeys the constraints set by quantum mechanics and by the De Broglie dispersion relation.

\section{Vlasov linearized solution}

The Vlasov equation describes the time evolution of the distribution function \( f(\vec{r}, \vec{\mathrm{v}}, t) \) for a collisionless plasma \cite{nicholson:plasma,Haas:2011}:
\begin{equation}
\frac{\partial f}{\partial t} + \vec{\mathrm{v}} \cdot \nabla_{\vec{r}} f + \frac{q}{m} \left( \vec{E} + \vec{\mathrm{v}} \times \vec{B} \right) \cdot \nabla_{\vec{\mathrm{v}}} f = 0,
\end{equation}
where \( q \) and \( m \) are the charge and mass of the particles, respectively, and \( \vec{E} \) and \( \vec{B} \) are the electric and magnetic fields. Although this equation is widely discussed in textbooks, we present its linearization procedure here not only to make the text self-contained, but also to establish consistent notation for the introduction of the Bohm potential in Sec.~\ref{Secvlaov}.

In the electrostatic case, the magnetic field is neglected (\( \vec {B} = 0 \)), and the electric field is derived from a scalar potential:

\begin{equation}
\vec {E} = -\nabla \phi(\mathrm{r}, t)
\end{equation}

Then, the Vlasov equation simplifies to

\begin{equation}
\frac{\partial f}{\partial t} + \vec{\mathrm{v}} \cdot \nabla_{\mathrm{r}} f - \frac{q}{m} \nabla \phi \cdot \nabla_{\mathrm{v}} f = 0.
\end{equation}
Assuming small perturbations around a stationary equilibrium:

\begin{eqnarray}
f(\mathrm{r}, \mathrm{v}, t) &=& f_0(\vec{\mathrm{v}}) + f_1(\vec {r}, \vec {\mathrm{v}}, t) \\ \nonumber \phi(\vec{r}, t) &=& \phi_1(\vec{r}, t)
\end{eqnarray}
with \( f_1 \ll f_0 \), and noting that \( f_0 \) is independent of time and position, the linearized Vlasov equation becomes:

\begin{equation}
\frac{\partial f_1}{\partial t} + \vec{\mathrm{v}} \cdot \nabla_{\mathrm{r}} f_1 - \frac{q}{m} \nabla \phi_1 \cdot \nabla_{\mathrm{v}} f_0 = 0.
\end{equation}
Using the Poisson equation
\begin{equation}
\nabla^2 \phi_1 = -\frac{q}{\varepsilon_0} \int f_1 \, d^3\mathrm{v},
\end{equation}
and plane wave perturbations
\begin{align}
f_1(\vec{r}, \vec{\mathrm{v}}, t) &= \hat{f}_1(\vec {k}, \vec{\mathrm{v}}, \omega) e^{i(\vec {k} \cdot \vec r - \omega t)} \\
\phi_1(\vec {r}, t) &= \hat{\phi}_1(\vec {k}, \omega) e^{i(\vec{k} \cdot \vec {r} - \omega t)}
\end{align}
we have
\begin{equation}
(-i \omega + i \vec k \cdot \vec {\mathrm{v}}) \hat{f}_1 = \frac{q}{m} (i \vec k \cdot \nabla_{\mathbf{v}} f_0) \hat{\phi}_1.
\end{equation}
from where  \( \hat{f}_1 \) is obtained :
\begin{equation}
\hat{f}_1 = -\frac{q}{m} \frac{{k} \cdot \nabla_{{{\mathrm{v}}}} f_0}{\omega - \vec {k} \cdot \vec {{\mathrm{v}}}} \hat{\phi_1}  \label{eq:f1}.
\end{equation}

The total density in the Fourier space is $\hat \rho(\vec k,\omega) = \int d^3v (f_0(\vec {\mathrm{v}}) + {\hat f}_1 (\vec k, \omega))$.
After an integration by parts and neglecting the surface term, we have
\begin{equation}
    {\hat f}_1 (\vec k, \omega) = \frac{q}{m} k^2 f_0(\vec {\mathrm{v}}) \frac{1}{(\omega-\vec k \cdot \vec {\mathrm{v}})^2  } \hat \phi_1(\vec k,\omega)
\label{fromVlasov}    
\end{equation}
or
\begin{eqnarray}
    {\hat \rho}_1 (\vec k, \omega) &=& \frac{q}{m}\left(\int d^3{\mathrm{v}}  \frac{ k^2 f_0(\vec {\mathrm{v}})}{(\omega-\vec k \cdot \vec {\mathrm{v}})^2  } \right) \hat \phi_1(\vec k) \\ \nonumber &=& K(\vec k, \omega)\hat \phi_1(\vec k,\omega)
\end{eqnarray}
where  ${\hat \rho}_1 (\vec k, \omega)$ is the Fourier transform of the induced density. We can invoke self-consistency  using the Poisson equation:
\begin{eqnarray}
     {\hat \rho}_1 (\vec k, \omega) &=&\frac{\epsilon_0 k^2}{q} \hat \phi_{1} (\vec k, \omega) \\ \nonumber &=& \frac{\epsilon_0k^2}{q}  \left(1-\epsilon (\vec k, \omega) \right)\hat \phi_T (\vec k, \omega)
     \label{poisson}
\end{eqnarray}
and therefore
\begin{equation}
    K(\vec k, \omega)\hat \phi_T(\vec k,\omega) = \frac{\epsilon_0k^2}{q}  \left(1-\epsilon (\vec k, \omega) \right)\hat \phi_T (\vec k, \omega)
\end{equation}
where $\phi_T (\vec k, \omega)$ is the self-consistent total potential (RPA framework)
and the dielectric function $\epsilon (\vec k, \omega) $ will read
\begin{eqnarray}
   \epsilon (\vec k, \omega) &=& 1-\frac{q}{\epsilon_0 k^2} K(\vec k, \omega)\\
   &=& = 1 - \frac{q^2}{\epsilon_0 m k^2} \int d^3\mathrm{v} \frac{k^2 f_0(\vec {\mathrm{v}}) }{(\omega-\vec k \cdot \vec {\mathrm{v}})^2 }.
\end{eqnarray}
For the case of a static electron gas ($f_0 = n \delta^3(\vec v)$) we have the well known Drude formula
\begin{equation}
       \epsilon (\vec k, \omega)  = 1 - \frac{\omega_p^2}{\omega^2}
\end{equation}
with $\omega_p^2 = \frac{q^2 n} {\epsilon_0 m}$.
\end{appendices}

\bibliographystyle{quantum}
\bibliography{bohm}
\end{document}